# Wave-front engineering by Huygens-Fresnel principle for nonlinear optical interactions in domain engineered structures

Yi-qiang Qin, Chao Zhang and Yong-yuan Zhu*

National Laboratory of Solid State Microstructures, Nanjing University,

Nanjing 210093, China

The wave-front engineering for nonlinear optical interactions was discussed. Using Huygens-Fresnel principle we developed a general theory and technique for domain engineering with conventional quasi-phase-matching structures being the special cases. By Fourier analysis we put forward the concept of local quasi-phase matching, which suggests that the quasi-phase matching is fulfilled only locally not globally. Experiments on focal effect of second-harmonic wave agreed well with the theoretical prediction. The proposed scheme combines three optical functions: generation, focusing and beam splitting of second-harmonic wave, thus making the device more compact. Further the proposed scheme can be used to perform the integration of multi-functional optical properties in nonlinear photonics, as well as expand the use of nonlinear optical devices.

* Corresponding author

The ferroelectric domain structure has been widely used for a variety of applications in both linear and nonlinear optics due to its multi-functional properties. In linear optics, it can be used for wave-front controlling by electrooptic effect with lens- or prism-like domain morphology (1-2). In nonlinear optics, spurred by the need for all-solid, short-wavelength and miniaturized laser devices, the study of quasi-phase-matching (QPM) has become a hot topic in the past two decades (3-12). QPM can be realized in a ferroelectric crystal by an artificial modulation of its second-order nonlinearity. The structure may be one-dimensional or two-dimensional (1D or 2D), periodic or quasi-periodic (13-20). Recently, a homocentrically poled $LiNbO_3$ and annular symmetry nonlinear frequency converters have been used to increase the angle acceptance and widen phase mismatch tolerance of second harmonic generation (SHG) (21-22). Parametric interactions in transversely patterned QPM gratings and even disordered domain structures have attracted considerable attention (23-25). In reference 24, the original 2D periodic arrays have been adjusted along the propagation direction of input fundamental wave, resulting in the relative phase shift between grating stripes. The classical Fraunhofer diffraction and the focusing of second harmonic (SH) beam in far-field have been demonstrated in such transversely patterned QPM gratings. The periodicity and QPM configuration still hold in the longitudinal direction for efficient SHG. Until now, basically, most of nonlinear parametric interactions realized in different domain engineered structures are totally treated in the framework of conventional QPM configuration. By introducing the concept of reciprocal space, the microstructured material can be viewed as a homogeneous medium except for the substitution of QPM for phase matching, where phase mismatch is compensated by reciprocal vectors. For QPM to be realized, the wave vectors (including the reciprocal vector) should all be well defined. However, if the reciprocal vectors as well as the wave vectors of the interaction waves are not well defined, the conventional QPM configuration is confronted with difficulties. Then would it be possible to develop a method with which nonlinear optical parametric interaction can be realized efficiently?

It is well known that Huygens-Fresnel principle plays an important role in classical optics. It states that every point on the primary wave-front acts as a source of secondary wavelets with the same frequency as the primary wave. These wavelets mutually interfere and the envelope of these wavelets gives rise to the wave-front at any later instant. Normally this process is performed in linear optics regime and in real space. In our present letter, we extend the Huygens-Fresnel principle to nonlinear optical parametric processes induced by second-order nonlinearity. That is, each point on the primary wave-front acts as a source of secondary wavelets of the fundamental as well as a source of, for example, the SH wave. As an example of proposed method, here the Huygens-Fresnel principle is used to design the domain structure, in which SH output can be controlled and focused into several points. The system acts as a complex lens for SH output, thus the proposed scheme may be also called wave-front engineering. Theoretical and experimental investigations on SHG wave-front engineering are reported. In order to explain the observed phenomenon, the concept of local QPM is put forward. It has shown that the local QPM configuration by Huygens-Fresnel principle can overcome the difficulties confronted either in structures without well defined reciprocal vector or in optical interacting waves without well defined wave vectors (such as for Gaussian or convergent beams).

In order to elucidate the above idea, we consider, as an example, the case of SHG in a domain engineered LiTaO$_3$ crystal. The input fundamental wave is a plane wave propagating along the x direction and the generated SH is focused into n points on the focal plane, all z polarized. Usually the domain pattern can be taken to be uniform in depth (the z direction); the system then simplifies to a 2D system, denoted as xy-plane. Within the plane the light propagation is isotropic. The multiple reflections, leading to photonic band gaps effects, are not present in this system due to the fact that the linear dielectric constant is constant in the whole structure. In addition, the fundamental wave keeps a plane wave when propagating in the structure. In such a 2D structure, the problem can be considered as scalar (26), which simplifies the notations. Under the slowly varying envelope approximation:

$$k^{2\omega} \cdot \nabla\left[E^{2\omega}(\vec{r})\right] \gg \nabla^2 E^{2\omega}(\vec{r}) \tag{1}$$

The evolution of the second harmonic amplitude can be written as a function of the pump field and the second-order coefficient $\chi^{(2)}$ (16):

$$k^{2\omega} \cdot \nabla\left[E^{2\omega}(\vec{r})\right] = -2i\frac{\omega^2}{c^2}\left[E^{\omega}(\vec{r})\right]^2 \chi^{(2)}(\vec{r})\exp\left\{-i\left[2k^{\omega}(\vec{r}) - k^{2\omega}(\vec{r})\right]\cdot \vec{r}\right\} \qquad (2)$$

Here $\mathbf{r} \equiv (x, y)$ is the 2D spatial coordinate. Generally, $E^{2\omega}$, $E^{\omega}$, $k^{2\omega}$ and $k^{\omega}$ are all spatial dependent. The nonlinear harmonic components can be described physically by Huygens-Fresnel principle. Because of nonlinearity in the atomic response, each atom develops an oscillating dipole moment which contains a component at frequency $2\omega$ and any material sample contains an enormous number of atomic diploes. Here the small part of crystal can be regarded as a point source which emits SH wave through stimulated dipoles oscillation. The initial phases of these radiations are determined by the phases of the incident fields, and modulated by the 2D domain engineered structure. In the case of focused SHG (either single focused SHG or multi-focused SHG), the wave vectors $k^{2\omega}(\vec{r})$ are spatial-dependent in the crystal. The SH wave focused at point ($X_i$, $Y_i$) propagated from point wave source at (x, y) with sample size *dxdy* is given:

$$dA_i^{2\omega} = -i\frac{1}{\sqrt{R_i(x,y)}} Kf(x,y)(A^{\omega})^2 \exp\left[-i\Delta k_i(x,y)\right]dxdy \qquad (3)$$

Here K is the coupling coefficient and f(x, y) is a 2D domain structural function. The distance between a point source and focused SHG point is $R_i(x,y) = \sqrt{(X_i - x)^2 + (Y_i - y)^2}$. In equation (3), $\frac{1}{\sqrt{R_i}}$ represents the decay law for amplitude of cylindrical wave and $A_j = \sqrt{\frac{n_j}{\omega_j}}E_j$ (j=$\omega$, 2$\omega$), respectively. It is noted that $\Delta k_i(x,y) = [2\vec{k}^{\omega}(r) - \vec{k}^{2\omega}(r)]\cdot \vec{r}$, the phase mismatch between the fundamental and harmonic waves, depends on the positions of both the point sources and the focused SHG points. For determination of structural function *f(x, y)*, we derived the following correlation function:

$$F(x,y) = \sum_{i=1,n} \frac{C_i}{\sqrt{R_i}} \exp\{-i[2\vec{k}^{\omega} \cdot \vec{r}(x,y) - \vec{k}^{2\omega} \cdot \vec{r}(x,y)]\} \qquad (4)$$

where $C_i$ is adjustable parameters; n is the number of multi-focused SHG points. Here the "correlation" means that the change of the coordinates of any one focused point $(X_i, Y_i)$ will change the value of F(x, y), even change its sign, and thus change the whole domain structure as can be seen below.

For the input being a fundamental plane wave, its phase function reduces to $k^{\omega}x$ and

$$\Delta k_i(x,y) = 2k^{\omega}x + k^{2\omega}R_i - k^{2\omega}(X_i \cos\theta_{1i} + Y_i \cos\theta_{2i}) \qquad (i=1, n) \qquad (5)$$

Here $k^{\omega}$ is the well defined wave vector of the fundamental; $\cos\theta_{1i}$ and $\cos\theta_{2i}$ are the directional cosine functions of $k^{2\omega}(\vec{r})$.

The structural function f(x, y) is determined by the correlation function:

$$f(x,y) = \begin{cases} 1, & \mathrm{Re}[F(x,y)] \geq 0 \\ -1, & \mathrm{Re}[F(x,y)] < 0 \end{cases} \qquad (6)$$

Integral over the whole system, the focused SHG at spatial points read:

$$A^{2\omega}(X_i, Y_i) = -K(A^{\omega})^2 \iint \frac{1}{\sqrt{R_i}} f(x,y) \exp[-i\Delta k_i(x,y)] dxdy \qquad (i=1,n) \qquad (7)$$

Formulas (1)-(7) constitute the basis of a general theory for domain engineering with conventional quasi-phase-matching structures being the special case. As an example, wave-front engineering for the multi-focused SHG by Huygens-Fresnel principle is discussed.

    Based on the theory introduced above, the 2D domain structures for the focused SHG are calculated. As a demonstration, Fig.1 shows the diagram of ferroelectric domain structures with the focused points designed to be on the exit surface of the crystal. Obviously the domain modulation is quite different from the conventional 1D and 2D structures. Fig. 1 (a) shows the domain structure for the single focused SHG. (X, Y) indicates the focused point. Near the focused point the domain boundary curves strongly; whereas far away from the focused point the domain morphology (as

shown in the inset) approximately approaches the conventional 1D periodic structure. It is interesting to note that the case investigated in Ref. [24] is just a special case when the focused point is far away from the parabolic domain pattern, where the periodicity holds approximately in the direction of input fundamental wave (shown in the inset of Fig. 1(a)).

Fig. 1(b) shows the domain structures for the dual focused SHG. The conventional stripe-like domains break into small bricks. The domain structures for the decal focused SHG with strange domain patterns are shown in the inset of Fig. 1(b). It can be seen that the more the focused points are, the more complex the domain structure is. Due to these domain structures, the SH wave-front generated from a fundamental plane wave is no longer a plane wave-front, rather becomes very complicated depending on the number of focused points, resulting in that the conventional QPM is out of work.

In order to verify the above scheme, we performed the experiments on focused SHG. The sample was fabricated by poling a 0.5mm thick z-cut $LiTaO_3$ single domain wafer at room temperature. The domain structure was designed such that the single focused SHG or dual focused SHG for the fundamental 1319nm z-polarized can be realized with the focused points located 10cm away from the exit surface. For dual focused SHG the two focused points are 2mm apart.

SHG of fabricated samples with 10mm length and 3mm width was tested using mode-locked Nd:YAG laser system pumped by CW diode laser with a pulse width of 150 ns and a repetition rate of 4 kHz. The fundamental wave was coupled into the polished end-face of the sample and propagated along the *x* axis of the sample. Three samples were tested in the experiments: periodic sample, samples with single focused SHG, and dual focused SHG, whose corresponding domain structures are exhibited in the bottom insets of Fig. 2, respectively.

Using weakly focused fundamental beam by a 15cm focal-length lens, the waist inside the sample is ~300μm where the intensity drops to $1/e^2$ of the maximum value and the confocal parameter for the system is $Z_0$~10cm much larger than the sample length 10mm. Thus the fundamental beam can be considered as a plane-wave. Figure 2 shows the SHG output CCD images for three different samples. Fig.2 (a) exhibits the SHG image in periodic sample, corresponding to the conventional 1D QPM scheme. The beam waist of red SHG is a little bit smaller (~ 80 %) than the waist of the input fundamental beam. Fig.2 (b) and 2(c) show the experimental results of the single and dual focused SHG (The calculated results are shown in the top insets). Fig.2 (b) shows the single focal SHG output. Obviously the considerable smaller beam waist of red SHG is obtained. Fig.2 (c) corresponds to the situation of dual focused SHG. By using brick-like domains (shown in the bottom inset of Fig.2 (c)) with neither translational symmetry nor annular symmetry, wave-fronts of SHG are controlled to be convergent towards the two focal points. In both cases, the measured SH minimum waists are about 120μm, agreeing well with the simulation result.

The SH powers are measured using a Si calibrated field master detector. In the experimental measurement the average SH output, with the fundamental power 260mW, are 110mW and 63mW in 1D periodic and brick-like structures, respectively. Here 42% SH conversion efficiency is achieved in periodic structure, 24% in brick-like structures. Experimental data also indicate almost the same SHG efficiency is obtained in 1D periodic structure and single focused SHG structure.

For deep understanding of the physical nature of wave-front engineering by Huygens-Fresnel principle, the Fourier spectra of the 2D domain structure for dual focused SHG (structure shown in the inset of Fig.2 (c)) are studied theoretically and experimentally. A He-Ne laser beam with wavelength 632.8nm is used to scan the z surface of the sample horizontally and vertically. The diffraction pattern is projected onto a screen and recorded by a CCD camera. Correspondingly Fourier transformation is performed numerically. From the results, some distinct features can be revealed. Along the horizontal direction the diffraction spectra have the same symmetry; whereas along the vertical direction the diffraction spectra show different symmetry. Fig.3 is the measured (a) and calculated (b) diffraction patterns obtained along the vertical direction, respectively. For the upper patterns, the most noticeable diffraction spots lay mainly in the second and forth quadrant. For the bottom ones, the most noticeable diffraction spots locate mainly in the first and third quadrant. They show mirror symmetric with the upper ones, which reflects the symmetry of domain pattern in real space. The symmetries of these patterns are totally different from the symmetry of the middle ones. This indicates the reciprocal vectors are spatial dependent. Actually, in the experiments, when moving the laser beam vertically from the upper part to the bottom part, the diffraction pattern has been observed to change gradually. That is, the symmetry of the pattern changes gradually from the top to the bottom: first the distribution of spots with quadrilateral-like inclined with respect to the central spot appears, then becomes quadrilateral totally, then changes to the triangle array in the middle parts. When further moving down the laser beam, the pattern changes inversely. The diffraction pattern changes slightly along the horizontal direction due to the fact that the focused points are designed rather far from the exit face of the sample. From the diffraction pattern exhibited above, it seems possible to define the so-called local QPM condition although the QPM condition could not be fulfilled globally. That is, the phase mismatch can be compensated locally with reciprocal vectors provided by local structure. In Fig. 3(b), we schematically show the reciprocal vectors used for local QPM, where $G_1$ is used for the upper focal point and $G_2$ for the lower one.

It is useful and interesting to compare the conventional QPM structure to current domain structure. When the single focused point of SHG is designed to be located infinitely, 2D modulated structure function f(x, y) degenerate into that of 1D periodic structure with the period of $2\pi/(k_{2\omega} - 2k_{\omega})$, corresponding to the conventional 1D QPM scheme. The domain structure can also degenerate into the conventional 1D structure when the focused points on the focal plane approach infinity. Recently research on the 2D QPM becomes a hot topic. In the framework of wave-front engineering with local QPM by Huygens-Fresnel principle, 2D periodic domain structure is a special case when multi-focused points of SHG are located infinitely at two perpendicular directions.

The advantage of the proposed scheme lies in its capability to perform several functions with a special designed domain structure. As one of the perspectives for application, the example presented above combines three functions: SHG, focusing and beam splitting, thus making the device more compact. The method can be also used for the tight focused Gaussian beam with which the efficient frequency conversion does not occur at phase-matching condition (27-28) and extended to other parametric processes such as down conversion. Studies on these processes by using local QPM are of potential interest in photonic applications. Thus Huygens-Fresnel principle may play an important role in nonlinear optical field with engineered domain structures.

We thank Dr. Xiao-peng Hu and Gang Zhao for help during this experiment. This work was supported by the State Key Program for Basic Research of China (Grant No. 2004CB619003); the National Natural Science Foundation of China (Grant Nos.10523001, 10504013 and 10674065).

Figure captions

Fig.1: The schematic diagram of ferroelectric domain structures for (a) single focused SHG (The inset is the domain morphology far from the focal point.), (b) dual focused SHG and (c) decal focused SHG, respectively. The focused points are indicated by ($X_i$, $Y_i$)

Fig.2: CCD images of SHG generated from (a) a periodic domain structure, (b) a domain structure for single focused SHG and (c) a domain structure for dual focused SHG, respectively. The bottom insets are their corresponding optical microscopic images of domain structures revealed by etching. The top insets in (b) and (c) are the calculated beam profiles of input fundamental wave and output SH wave at focusing plane. The solid and dash lines correspond to input fundamental beam and output SH wave, respectively.

Fig.3: Fourier spectra of domain structure shown in the inset of Fig. 2(c) for dual focused SHG obtained along the vertical direction: (a) the experimental and (b) the calculated results. The schematic diagrams of local quasi-phase-matching are indicated.

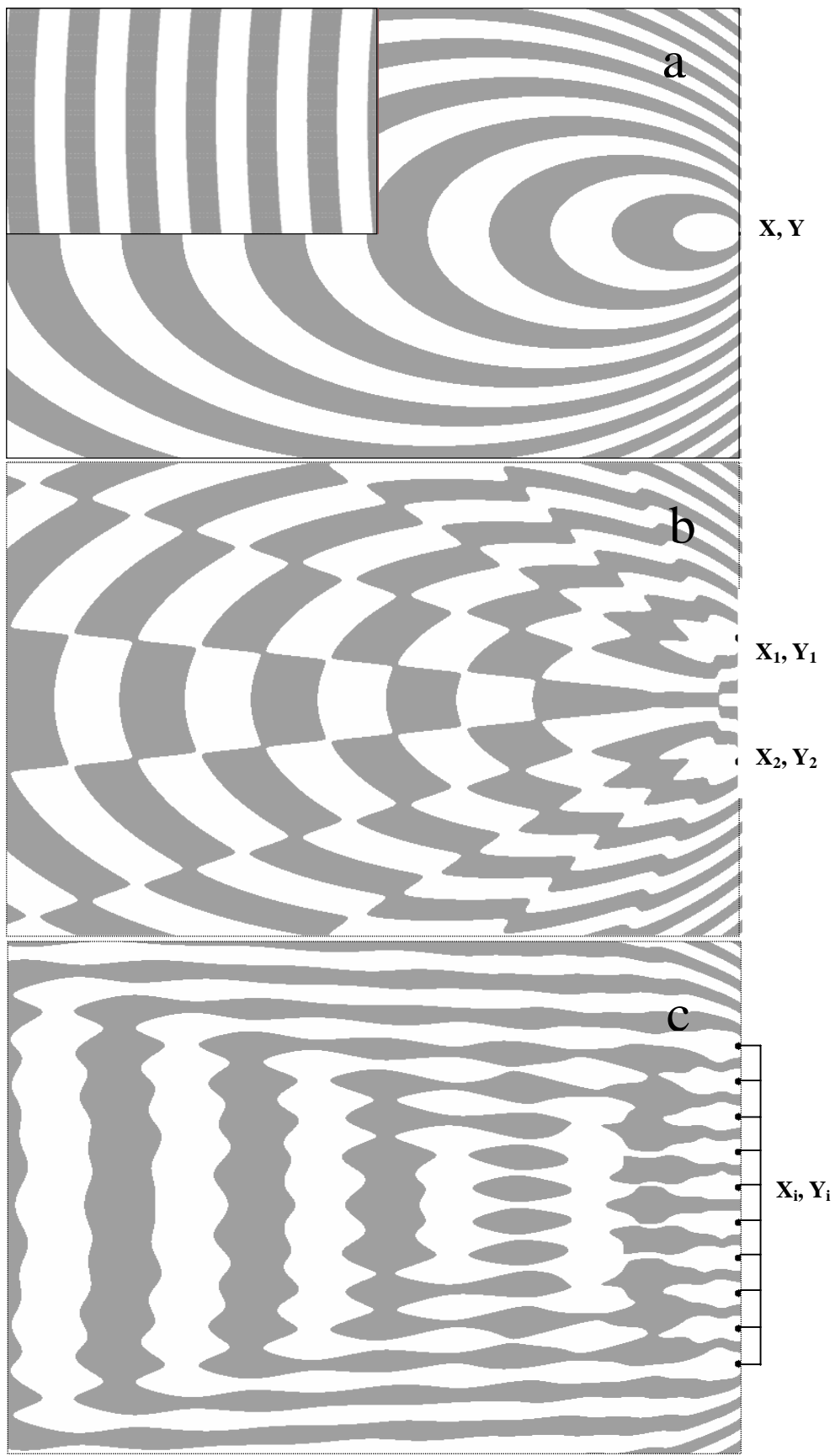

Figure 1

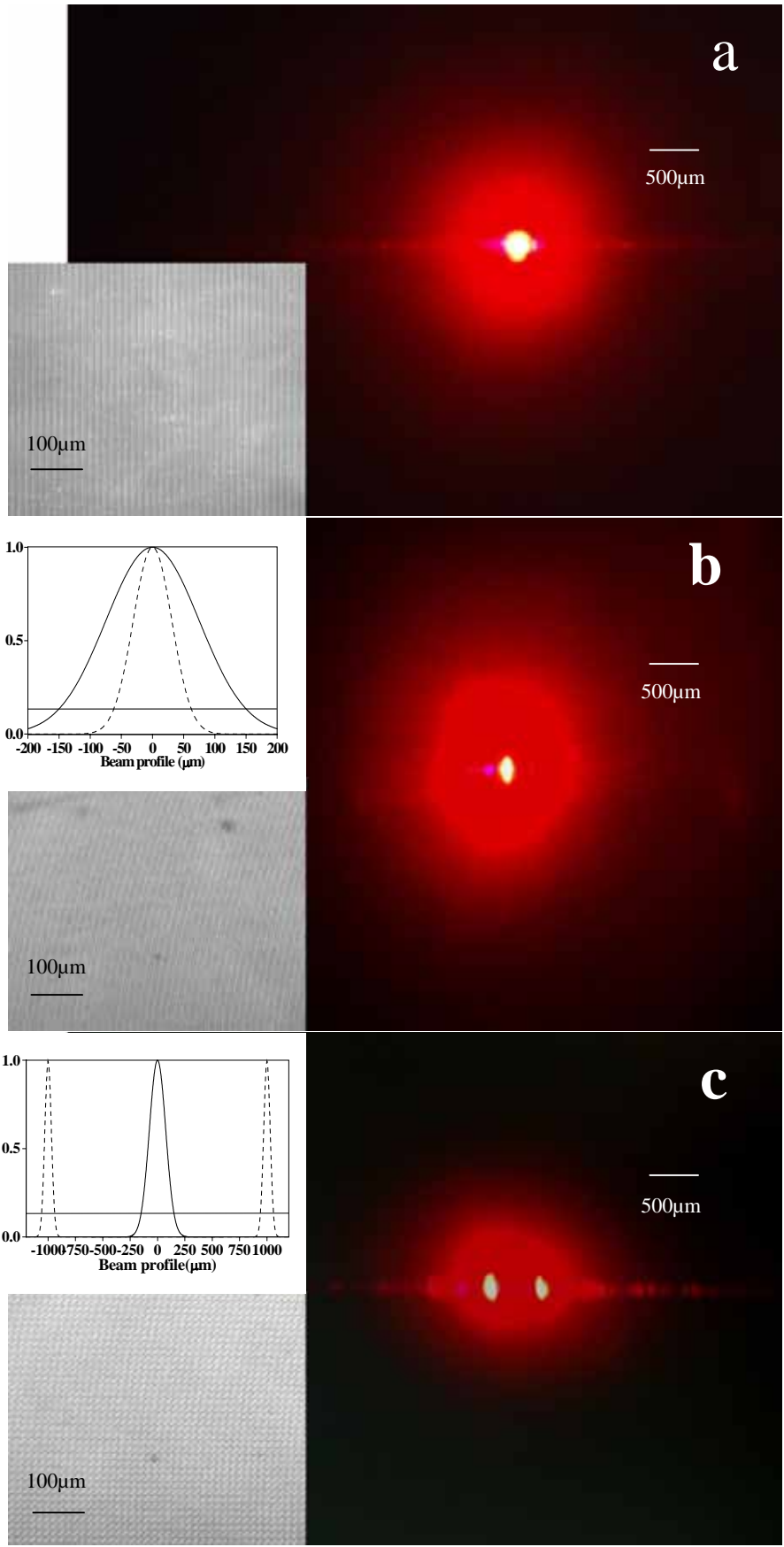

Figure 2

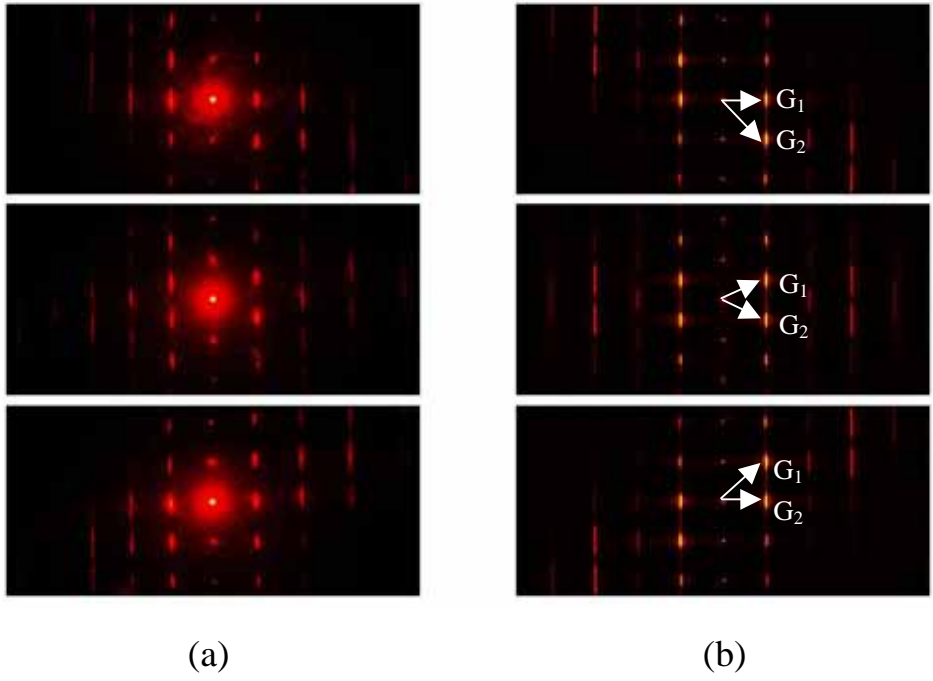

(a)          (b)

Figure 3